\documentclass[twocolumn,prd,superscriptaddress,showpacs,floatfix,%
preprintnumbers, nofootinbib,hyperref]{revtex4}  
\usepackage{epsfig}   
\usepackage{bm}  
\usepackage{color}  
\usepackage[usenames,dvipsnames]{xcolor}

\newcommand{\ben}{\begin{equation}}
\newcommand{\een}{\end{equation}}

\newcommand{\gtrsim}{\,\rlap{\lower3.7pt\hbox{$\mathchar\sim$}}
\raise1pt\hbox{$>$}\,}
\newcommand{\lesssim}{\,\rlap{\lower3.7pt\hbox{$\mathchar\sim$}}
\raise1pt\hbox{$<$}\,}

\newcommand{\be}{\begin{equation}}
\newcommand{\ee}{\end{equation}}  
\newcommand{\bea}{\begin{eqnarray}}
\newcommand{\eea}{\end{eqnarray}}

\def\theta{\vartheta}   

\begin{document}

\preprint{LAPTH-037/12}

\title{{\color{Red}Flavor stability analysis of dense supernova neutrinos with \\ 
 flavor-dependent angular distributions}}

\author{Alessandro Mirizzi} 
\affiliation{II Institut f\"ur Theoretische Physik, Universit\"at Hamburg,
Luruper Chaussee 149, 22761 Hamburg, Germany} 

\author{Pasquale Dario Serpico} 
\affiliation{LAPTh, Univ. de Savoie, CNRS, B.P.110, Annecy-le-Vieux F-74941, France}


\begin{abstract}
Numerical simulations of the supernova (SN) neutrino self-induced flavor conversions, associated with the neutrino-neutrino interactions 
in the deepest stellar regions, have been typically carried out assuming the ``bulb-model''. In this approximation, neutrinos are taken to 
be emitted half-isotropically by a common neutrinosphere. In the recent Ref.~\cite{Mirizzi:2011tu}  we have   removed this assumption by introducing
 flavor-dependent angular distributions for SN neutrinos, as suggested by core-collapse simulations. We have found that in this 
case  a novel multi-angle instability in the self-induced flavor transitions can arise. In this work we perform an extensive study of  this  effect, carrying out  a linearized  flavor stability analysis   for different
SN neutrino energy fluxes and angular distributions, in both normal and inverted neutrino mass hierarchy. We confirm that spectra of different $\nu$ species which cross in angular space (where $F_{\nu_e}=F_{\nu_x}$ and $F_{\bar\nu_e}=F_{\bar\nu_x}$) present a significant enhancement of the flavor instability, and a shift of the onset of the flavor conversions at smaller radii with respect to the case of an isotropic neutrino emission. We also illustrate how a qualitative (and sometimes quantitative) understanding
of the dynamics of these systems follows from a stability analysis.
\end{abstract}


\maketitle

\section{Introduction} 
Flavor conversions of neutrinos emitted  from  core-collapse supernovae (SNe)
represent a diagnostic tool to get crucial information about their  mixing parameters
and the SN dynamics in the deepest stellar regions~\cite{Raffelt:2010zza}.
In particular, SNe are a unique laboratory to probe neutrino oscillations in extreme conditions.
Supernova  neutrinos can  interact not only with the stellar medium
via the Mikheyev-Smirnov-Wolfenstein (MSW) effect~\cite{Matt,Dighe:1999bi},  but also with other
neutrinos  ($\nu$) and  antineutrinos 
(${\overline\nu}$) as well. 
It was pointed out that large $\nu$ densities in the deepest stellar regions can result
in significant coherent $\nu$--$\nu$ forward
scatterings~\cite{Pantaleone:1992eq,Qian:1994wh}. A few years ago it was discovered
that $\nu$ self-interactions can
give rise to collective $\nu$ flavor oscillations inside the
SN~\cite{Duan:2006an,Hannestad:2006nj,Pehlivan:2011hp}
(see~\cite{Duan:2010bg} for a recent review).
The most important observational consequence of this collective behavior is a
swap of the $\nu_e$ and ${\overline\nu}_e$ spectra with the non-electron $\nu_x$
and ${\overline \nu}_x$ spectra in certain energy
ranges~\cite{Fogli:2007bk,Dasgupta:2009mg}. It has been argued that ``spectral splits'' at the edge of each swap interval
would be observable in the high-statistics $\nu$ signal from the next galactic SN, 
allowing to get crucial information about the unknown $\nu$ mass ordering 
(see, e.g.,~\cite{Duan:2007bt}).

Self-induced oscillation  effects crucially depend  
 on the inner  boundary conditions fixed
for the further flavor evolution. Indeed, these non-linear
 flavor conversions are associated 
to instabilities in the flavor space, that develop around the crossing points 
in the energy spectra of the different $\nu$ species (where $F_{\nu_e}=F_{\nu_x}$ and 
$F_{{\bar\nu}_e}=F_{{\bar\nu}_x}$)~\cite{Dasgupta:2009mg}. 
The number and the position of the crossing points  depend on the
ordering of the original SN $\nu$ fluxes. Since this latter can   change during the different
post-bounce stages, significant temporal variations are expected in the 
pattern of the spectral splits~\cite{Dasgupta:2009mg,Duan:2010bf,Mirizzi:2010uz}.   
An important layer of complication in this description is associated to the 
 the current-current nature of the $\nu$-$\nu$  weak-interaction
Hamiltonian. This implies that the interaction
energy between neutrinos of momenta ${\bf p}$ and ${\bf q}$ is
proportional to $(1 - {\bf v_p} \cdot {\bf v_q})$, where ${\bf v_p}$ is the neutrino
velocity~\cite{Pantaleone:1992eq,Qian:1994wh,Sawyer:2008zs}. In a non-isotropic medium this velocity-dependent
term would not average to zero, producing
a different refractive index for neutrinos propagating on
different trajectories. This is the origin of the so-called
``multi-angle effects,'' which in some case
can dramatically affect the development of the self-induced flavor conversions, 
producing a quick flavor decoherence~\cite{Raffelt:2007yz,EstebanPretel:2007ec}  or suppressing flavor conversions otherwise possible 
for an   a isotropic neutrino emission~\cite{Duan:2010bf,Mirizzi:2010uz}.
Moreover, it has been shown that the presence of the ordinary matter background 
would cause a multi-angle suppression of the collective oscillations, when the matter density
dominates over the neutrino one~\cite{EstebanPretel:2008ni}. This situation is expected to occur during the early times accretion phase~\cite{Chakraborty:2011nf,Chakraborty:2011gd,Sarikas:2011am,Saviano:2012yh,Sarikas:2012vb}.
 
The characterization of multi-angle effects is then a key-ingredient to obtain a realistic 
description of the self-induced neutrino flavor conversions. 
In this context, it is expected that the $\nu$ angular distributions at emission would 
play an important role in determining the $\nu$-$\nu$ interaction strength.  
Until recently 
 numerical simulations for the SN $\nu$ flavor conversions  have 
been based on the so-called ``bulb model'' (see, e.g,~\cite{Duan:2006an,Fogli:2007bk}), 
where $\nu$'s of different species are considered as 
emitted ``half-isotropically'' (i.e. with  all outward-moving angular modes 
equally occupied and all the backward-moving modes
empty) by a common spherical ``neutrinosphere,''
in analogy with a blackbody emission. 
However,  realistic supernova simulations show
that $\nu$  angular distributions at decoupling are far from being half-isotropic and, above all, are flavor-dependent
(see, e.g.,~\cite{Sarikas:2011am,Ott:2008jb,Dasgupta:2011jf}).
The presence of non-trivial  angular distributions was claimed in~\cite{Sawyer:2005jk}  to produce 
a novel multi-angle instability in the self-induced flavor evolution of a toy model of $\nu$ gas. 
However, that conclusion was based on an analysis performed with a small number of angular modes, making
challenging to rely on this claim to infer conclusions for  the realistic SN $\nu$ case 
 without a dedicated large-scale numerical study. Triggered also by this warning,
 in~\cite{Mirizzi:2011tu} we performed numerical simulations of the self-induced 
flavor conversions with non-trival $\nu$ angular distributions, finding 
remarkable effects on the flavor evolution. 
In particular, when flavor-dependent angular distributions  lead to crossing points in the angular spectra of different $\nu$ species,  a new multi-angle instability can develop, in analogy to
the known instability triggered by crossing points  in the energy domain.
We find cases in which this  multi-angle instability can  shift the onset of the flavor conversions toward  low-radii and  produce
a smearing of the splitting features found with trivial $\nu$ emission models.
In order to achieve a semi-analytical understanding of the phenomenon, in that Letter we proposed to apply to our problem the linearized stability analysis
recently worked out in~\cite{Banerjee:2011fj}. By seeking
an exponentially growing solution in the eigenvalue equations associated with the linearized equations of motion for
the neutrino ensemble, this method allowed us to determine the onset of the flavor conversions. In a specific scenario, we compared the growth of this mode in the half-isotropic case with different cases with non-trivial $\nu$ angular distributions, finding a significant enhancement of the instability in these latter cases.

The purpose of this follow-up work is to take a closer look to the multi-angle instability, triggered 
by non-trivial angular distributions. In particular we aim to use the stability analysis to perform an investigation 
on the dependence of this effect on the initial SN $\nu$ flux ordering, and on the $\nu$ mass hierarchy.
Here is the plan of our work. In Sec.~\ref{setup} we  introduce the setup for the flavor-stability analysis, 
describing the non-linear equations for the $\nu$ flavor  evolution in SNe, and the consistency equations coming
from their linearization. 
In Sec.~\ref{stability} we present our models for the supernova neutrino emission
for cases with a different flux ordering and  angular distributions. For these different cases we show  the results of the stability analysis in the two neutrino mass hierarchies. Finally in Sec.~\ref{conclusions} we comment on our results and we conclude. 

\section{Setup of the stability analysis}\label{setup}

\subsection{Equations of motion}

Our main goal is to perform a numerical stability analysis for the cases  studied  in~\cite{Mirizzi:2011tu}.
Therefore, we use the same setup followed in that paper. 
In particular, we work in  a two-flavor oscillation scenario, associated to the 
atmospheric mass-square difference $\Delta m^2_{\rm atm}= 2 \times 10^{-3}$~eV$^2$ and
and with the small (matter suppressed) in-medium mixing $\Theta = 10^{-3}$.  
Since we aim at isolating the effect of the multi-angle instability, associated with $\nu$ non-trivial angular
distributions,
we assume in the following that self-induced flavor conversions are not matter suppressed (as expected instead at $t \lesssim 1$~s
after the core bounce), ignoring the ordinary matter background  in the equations of motion. 
Therefore, also when we will consider neutrino spectra representative of the accretion phase
(at $t \lesssim 0.5$~s
after the core bounce) we will neglect the matter term in order to isolate only the  effect
of the angular instability. However, it has been found in~\cite{Sarikas:2011am,Saviano:2012yh}
that taking realistic angular distributions during the accretion phase and a dominant matter term,
self-induced flavor conversions would be completely inhibited.
Conversely, we checked that the sub leading matter term during the cooling phase
(at $t \gtrsim 1$~s
after the core bounce) would have a minor 
impact on the flavor instability associated with the non-trivial angular distributions.     

If we denote with $\theta_r$ the angle of a given neutrino trajectory with respect to the radial direction, flux conservation implies that that it will depend on the radial coordinate $r$ even for straight line propagation (as we consider here). 
It is thus convenient to parameterize every angular mode in terms of its emission angle $\theta_R$ relative to the radial direction
of the neutrinosphere, that here we schematically fix at $R=10$~km.  
For simplicity, we neglect possible residual scatterings that could affect $\nu$'s after the 
neutrinosphere, producing a small ``neutrino halo'' that would broaden the $\nu$ angular 
distributions~\cite{Cherry:2012zw}.
Indeed, it is expected that this effect may be relevant only at early times, where self-induced conversions
would  be matter suppressed~\cite{Sarikas:2012vb}.
For
a half-isotropic distribution the occupation numbers are distributed as $d n/d \cos \theta_R = const.$,
or equivalently the radial fluxes are distributed as $d \Phi/ d \cos \theta_R \propto \cos\theta_R$~\cite{EstebanPretel:2007ec}.
A further simplification is obtained if one labels the different angular modes in terms of the variable $u=\sin^2\theta_R$,
as in~\cite{EstebanPretel:2007ec,Banerjee:2011fj}. Note that 
for an half-isotropic emission at the neutrinosphere  the $\nu$ angular distribution of the radial fluxes
 is a box spectrum in $0\leq u \leq 1$, since $ d \cos \theta_R/du\propto (\cos\theta_R)^{-1}$ which cancels the $\cos\theta_R$
 dependence previously mentioned.  At a radius $r$, the radial velocity of a mode with angular label $u$ is
$v_{u,r} = (1-u R^2/r^2)^{1/2}$~\cite{EstebanPretel:2007ec}.
Following~\cite{Banerjee:2011fj}, we write the equations of motion for the flux matrices  $\Phi_{E,u}$  as function of the radial coordinate.
The diagonal $\Phi_{E,u}$ elements are
the ordinary number fluxes $F_{\nu_{\alpha}}(E,u)$ 
integrated
over a sphere of radius $r$.
We normalize the flux matrices to the total ${\overline\nu}_e$ number flux $N_{{\bar\nu}_e}$ at the neutrinosphere.
 Conventionally, we use negative $E$ and negative
number 
fluxes for anti-neutrinos. The off-diagonal elements,
which are initially zero, carry a phase information due to 
flavor mixing.
Then, the equations of motion read~\cite{Banerjee:2011fj,Sigl:1992fn}
\begin{equation}
\textrm{i}\partial_r \Phi_{E,u}=[H_{E,u},\Phi_{E,u}] \,\ 
\label{eq:eom1}
\end{equation}
with the Hamiltonian~\cite{Pantaleone:1992eq,Qian:1994wh,Banerjee:2011fj,Sigl:1992fn}
 \begin{eqnarray}
& & H_{E,u} = \frac{1}{v_{u}}\frac{M^2}{2E} \nonumber \\
 &+& \frac{\sqrt{2}G_F}{4\pi r^2}\int_{-\infty}^{+\infty}d E^\prime
 \int_{0}^{1}du^\prime \left(\frac{1-v_{u}v_{u^\prime}}{v_{u}v_{u^\prime}}
 \right)\Phi_{E^\prime,u^\prime} \,\,
 \label{eq:eom2}
 \end{eqnarray}
 in which we neglected matter effects.
 The matrix $M^2$ of neutrino mass-squares causes vacuum
flavor oscillations and the term at second lines represents the $\nu$-$\nu$ refractive 
term. 
In particular, the 
 factor proportional to the neutrino velocity $v_{u,r}$ in 
the $\nu$-$\nu$ interaction term
implies ``multi-angle'' effects for neutrinos moving on different trajectories~\cite{Pantaleone:1992eq,Qian:1994wh, Duan:2006an,Sigl:1992fn}. In order to properly simulate numerically this effect one needs 
to  follow a large number
$[{\mathcal O}(10^3)]$ of interacting $\nu$ modes.


\subsection{Stability conditions}
In order to perform the stability analysis we closely follow the
prescriptions presented in~\cite{Banerjee:2011fj} and summarized in the following. 
Firstly, we switch to the 
frequency variable $\omega= \Delta m^2_{\rm atm}/2E$ so that $E(\omega)=|\Delta m^2_{\rm atm}/2\omega|$ and  we introduce the 
neutrino flux difference distributions $g_{\omega,u}\equiv g(\omega,u)$ defined as
\begin{eqnarray}
&&g_{\omega,u}=\frac{|\Delta m_{\rm atm}^2|}{2\omega^2}\times\bigg\{\Theta(\omega)\left[F_{\nu_e}(E(\omega),u)-F_{\nu_x}(E(\omega),u\right)]\nonumber\\
&&+\Theta(-\omega)\left[F_{\nu_x}(E(\omega),u)-F_{\overline\nu_e}(E(\omega),u)\right]\bigg\}\,,
\end{eqnarray}
normalized  to the total ${\overline\nu}_e$ flux at the neutrinosphere. Whenever referring to numerical values for the variables $\omega,\mu$ we shall implicitly quote them in km$^{-1}$, as appropriate for the SN case.  Note that  $g_{\omega,u}$ is defined also for negative $\omega$, where it represents the difference of fluxes in the antineutrino sector in the opposite ordering. 
Then, we   write the flux matrices in the form~\cite{Banerjee:2011fj}
\begin{equation}
\Phi_{\omega,u}= \frac{\textrm{Tr}\,\Phi_{\omega,u}}{2}+
\frac{g_{\omega,u}}{2}
\left( \begin{array}{cc} s_{\omega,u} &  S_{\omega,u} \\
S^{\ast}_{\omega,u} & -s_{\omega,u} 
\end{array} \right) \,\ ,
\end{equation}
where $\textrm{Tr}\,\Phi_{\omega,u}$ is conserved and then irrelevant for the flavor conversions, and the initial conditions for the  
``swapping matrix'' in the second term on the right-hand side are $s_{\omega,u}=1$ and $S_{\omega,u}=0$.
Self-induced flavor transitions  start when the off-diagonal term $S _{\omega,u}$
 exponentially grows.

In the small-amplitude limit $|S _{\omega,u}|\ll 1$, 
and at far distances from the neutrinosphere $r \gg R$,
the linearized evolution equations for  $S _{\omega,u}$
in inverted mass hierarchy (IH, $\Delta m^2_{\rm atm}<0$)
assume the form~\cite{Banerjee:2011fj}
\begin{eqnarray}
\textrm{i}\partial_r S _{\omega,u} &=& (\omega + u\epsilon \mu) S _{\omega,u} \nonumber \\
&-&\mu \int du^\prime d\omega^\prime (u+u^\prime)g_{\omega^\prime, u^\prime} S _{\omega^\prime,u^\prime} \,\ ,
\label{eq:stab}
\end{eqnarray}
where 
\begin{equation}
\epsilon = \int du \,\ d\omega \,\ g_{\omega,u} \,\ ,
\label{eq:asy}
\end{equation}
quantifies the ``asymmetry'' of the neutrino spectrum, normalized to the total ${\overline\nu}_e$ 
number flux.
The $\nu$-$\nu$ interaction strength is given by
\begin{eqnarray}
\mu &=& \frac{\sqrt{2}G_F N_{{\bar\nu}_e}}{4 \pi r^2}\frac{R^2}{2 r^2} \nonumber \\
&=& \frac{3.5 \times 10^{5}}{r^4} \left(\frac{L_{{\overline\nu}_e}}{10^{52} \,\ \textrm{erg/s}}\right)
\left(\frac{15 \,\ \textrm{MeV}}{\langle E_{{\overline\nu}_e} \rangle}\right) 
 \left(\frac{R}{10 \,\ \textrm{km}} \right)^2 . \nonumber
\end{eqnarray}
One can write the solution of the linear differential equation [Eq.~(\ref{eq:stab})]
in the form $S _{\omega,u} = Q _{\omega,u} e^{-i\Omega r}$ with complex frequency
$\Omega= \gamma + i \kappa$ and eigenvector $Q _{\omega,u}$. A solution with 
$\kappa >0$ would indicate an exponential increasing $S _{\omega,u}$, i.e. an instability.
The solution of Eq.~(\ref{eq:stab}) can then be recast in the form of an eigenvalue equation
for $Q _{\omega,u}$. Splitting this equation into its real and imaginary parts one arrives
at two real equations that have to be satisfied~\cite{Banerjee:2011fj}
\begin{eqnarray}
(J_1-\mu^{-1})^2 &=& K_1^2 + J_0 J_2 -K_0 K_2 \,\ , \nonumber \\
(J_1-\mu^{-1}) &=& \frac{J_0 K_2 + K_0 J_2}{2 K_1} \,\ ,
\label{eq:JnKn}
\end{eqnarray}
where
\begin{eqnarray}
J_n &=& \int d\omega \,du\, g_{\omega,u} u^n \frac{\omega -\omega_{\rm res}}
{(\omega -\omega_{\rm res})^2 + \kappa^2} \,\ , \nonumber \\
K_n &=& \int d\omega \, du\, g_{\omega,u} u^n \frac{\kappa}
{(\omega -\omega_{\rm res})^2 + \kappa^2} \,\ ,
\label{eq:consit}
\end{eqnarray}
and we introduced the \emph{resonant frequency}
\begin{equation}
\omega_{\rm res}(\gamma)= \gamma - u\epsilon \mu  \,\ .
\label{eq:res}
\end{equation}

A flavor instability is present whenever Eqs.~(\ref{eq:JnKn}) admit a  solution $(\gamma, \kappa)$. 
When an instability occurs, for a given  angular mode $u_0$ the function 
$|Q _{\omega,u_0}|$ is a Lorentzian~\cite{Dasgupta:2009mg}, centered around $\omega_{\rm res}$,  with a width $\kappa$. 
Finally, we remind that the consistency equations in the normal mass hierarchy (NH) case 
($\Delta m^2_{\rm atm}>0$) are obtained simply changing the sign of $g_{\omega, u}$
in Eqs.~(\ref{eq:consit})~\cite{Banerjee:2011fj}.

\section{Stability analysis for different models}\label{stability}

\subsection{Models for supernova neutrino emission}

\begin{figure}  
\includegraphics[angle=0,width=0.7\columnwidth]{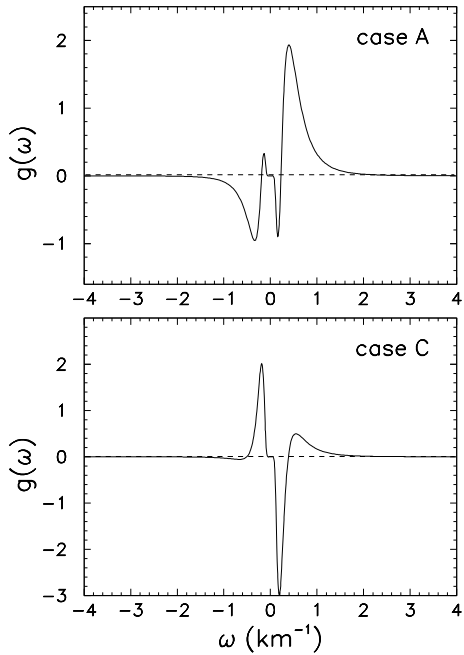}
\caption{Spectrum $g(\omega)$ for the case  $\cal A$ (upper panel) and $\cal C$
(lower panel).
\label{fig1}} 
\end{figure}  

In order to perform our stability analysis we have to fix a neutrino emission model. 
Since our main goal is to explore in more detail the numerical findings presented in~\cite{Mirizzi:2011tu}, in the following we refer
to the cases discussed in that work. 
We remark that energy and angular distributions of SN $\nu$'s  entering $F_{\nu_\alpha}(E,u)$ are not independent
of each other. However, schematically we assume that the angular distributions are energy
independent. Then, we can factorize the $\nu$ flux of each flavor as  
$F_{\nu_\alpha}(E,u)= N_{\nu_\alpha} \times \varphi_{\nu_\alpha}(E) \times
U_{\nu_\alpha}(u).$
The $\nu$ number $N_{\nu_\alpha}=L_{\nu_\alpha}/{\langle E_{\nu_\alpha}\rangle}$ is expressed in terms of
the $\nu$ luminosity  $L_{\nu_\alpha}$ and of  the $\nu$ average energy ${\langle E_{\nu_\alpha}\rangle}$ of the different species.
The function $\varphi_{\nu_\alpha} (E)$ is the normalized $\nu$ energy spectrum ($\int dE \varphi_{\nu_\alpha} (E)=1$)
and $U_{\nu_\alpha}(u)$ is the normalized angular distribution ($\int du U_{\nu_\alpha}(u)$=1). 

We parametrize the energy spectrum  as in~Ref.~\cite{Keil:2002in}
\begin{equation}
\varphi(E)=\frac{(1+\alpha)^{1+\alpha}}{\Gamma(1+\alpha)}\frac{E^\alpha}{\langle E_\nu\rangle^{\alpha+1}}\exp\left[-\frac{(1+\alpha)\,E}{\langle
E_\nu\rangle}\right]\, ,
 \label{eq:varphi}
\end{equation}
where we fix the spectral parameter to  $\alpha=3$ for all species.
Following~\cite{Mirizzi:2011tu}, we fix the neutrino average energies at
\begin{equation}
(\langle E_{\nu_e}\rangle, \langle E_{{\bar\nu}_e}\rangle, \langle E_{\nu_x}\rangle)=(12, 15, 18)\,\  
\textrm{MeV} \,\ .
\end{equation}
Concerning the possible $\nu$ flux ordering we consider two cases.  As representative of the accretion phase (labelled as case ${\cal A}$), we take
\begin{equation}
N_{\nu_e}:N_{\bar\nu_e}:N_{\nu_x}=1.50:1.00:0.62 \,\ ,
\end{equation}
corresponding to an asymmetry paramter $\epsilon = 0.50$ in Eq.~(\ref{eq:asy}).
Instead, as representative of the cooling phase, dubbed  case ${\cal C}$,
we choose
\begin{equation}
N_{\nu_e}:N_{\bar\nu_e}:N_{\nu_x}=1.13:1.00:1.33  \,\ ,
\end{equation}
giving $\epsilon = 0.13$.

In Figure~1  we show the function $g_{\omega}= \int du\,  g_{\omega,u}$ for the case
${\cal A}$ (upper panel) and for the case ${\cal C}$ (lower panel). 
Both spectra present three crossing points, where  
 $g_{\omega}=0$, so naively one would expect similar instability conditions in these two cases. However, as we will discuss in the following, the instabilities in these two cases are significantly different.

\begin{figure}[!t]  
\includegraphics[angle=0,width=1.\columnwidth]{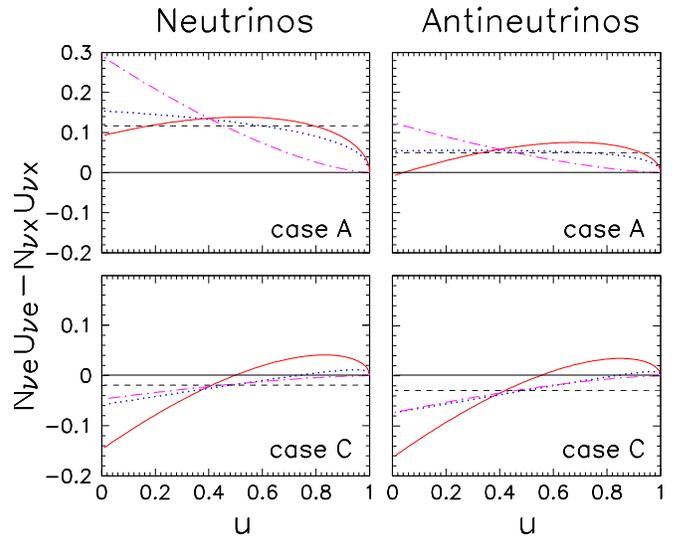}
  \caption{Difference of the energy-integrated angular spectra $N_{\nu_e} U_{\nu_e}(u) -
N_{\nu_x} U_{\nu_x}(u)$ for $\nu$'s (left panels) and $\overline\nu$'s (right panels)
for the case ${\cal A}$ (upper panels) and ${\cal C}$ (lower panels).
 The different curves correspond to $\beta_e=\beta_x=0.0$ (dashed curves), $\beta_e=1.0, \beta_x=1.5$
 (dotted curves), $\beta_e=1.0, \beta_x=3.0$ (continuous curves), $\beta_e=3.0, \beta_x=3.0$
(dash-dotted curves). 
\label{fig2}} 
\end{figure}  

Concerning the neutrino angular distributions, 
we  use 
the simple toy model introduced in~\cite{Mirizzi:2011tu} to capture the main deviations with respect to the half-isotropic bulb model, 
where $U_{\nu_\alpha}= 1$ for all the $\nu$ species. In particular,  we choose forward-peaked distributions
$U_{\nu_\alpha}(u) \propto (1-u)^{\beta_\alpha/2}$. For simplicity in the following  we  assume $U_{\nu_e}=U_{{\bar\nu}_e}$ and we  refer to these cases:
\begin{itemize}
\item[-] $(\beta_e=\beta_x=0.0)$ \,\ ,
\item[-] $(\beta_e=1.0 \,,\, \beta_x=1.5)$ \,\ ,
\item[-] $(\beta_e=1.0 \,,\, \beta_x=3.0)$ \,\ ,
\item[-] $(\beta_e=3.0 \,,\, \beta_x=3.0)$ \,\ .
\end{itemize}

In Figure~\ref{fig2} we we plot the difference $N_{\nu_e}U_{\nu_e}(u)-N_{\nu_x}U_{\nu_x}(u)$ for $\nu$'s (left panels) and the analogous one for $\overline\nu$'s (right panels) for the case ${\cal A}$ (upper panels) and for the case 
 ${\cal C}$ (lower panels) for the four $(\beta_e,\beta_x)$ cases.
 The crossing points are given by the intersection of these functions with the horizontal lines at zero.   
In the  half-isotropic case $\beta_e=\beta_x=0.0$, the differences of angular spectra do not present any crossing point in the angular
variable, while in all the other cases, the spectra present a crossing point at $u=1$ where $N_{\nu_e}U_{\nu_e}(u)=N_{\nu_x}U_{\nu_x}(u)$, and all the angular distributions $U_{\nu_\alpha}$ vanish.
 In the case $\cal A$ the angular distributions do not present other crossing points at finite $u$.  
Also for the $\overline\nu$'s  in the $\beta_e=1.0$, $\beta_x=3.0$ case,  the crossing point at finite $u$ is  so close to the edge of the spectrum at $u=0$, that cannot be distinguished by it.
Instead,  in the case $\cal C$  the (anti)neutrino angular spectra present 
also one crossing at finite  $0<u<1$.
In~\cite{Mirizzi:2011tu}, we associated the speed-up of the multi-angle instability  with the presence or absence of crossing points
 in the energy-integrated angular spectra.
 In the following, we shall confirm this explanation by performing an extensive stability analysis the different cases.

\subsection{Case ${\cal A}$}

The $g_{\omega}$ function for the case ${\cal A}$ (Fig.~\ref{fig1} upper panel),
 presents three crossing points in the $\omega$ variable.
According to what explained in~\cite{Dasgupta:2009mg} and confirmed by the stability analysis in~\cite{Banerjee:2011fj}, 
the instabilities that trigger self-induced oscillations are associated to crossings
with positive slope in IH  and to 
crossings with negative slope  in NH. 
Therefore, one would have expected in this case a single self-induced spectral swap around $\omega=0$
in normal  hierarchy, and  two spectral swaps in IH around the two crossings at finite $\omega$. 
However, as observed in many numerical simulations for flux ordering with 
$N_{\nu_e}>N_{\bar\nu_e}>N_{\nu_x}$, the flavor evolution 
presents only a broad swap around $\omega=0$ in IH and no conversion in NH. 
 Indeed, as discussed in~\cite{Dasgupta:2009mg}, a narrowly spaced triple crossing
can superficially act like a single one at $\omega=0$.

\begin{figure}[!t]  
\includegraphics[angle=0,width=0.8\columnwidth]{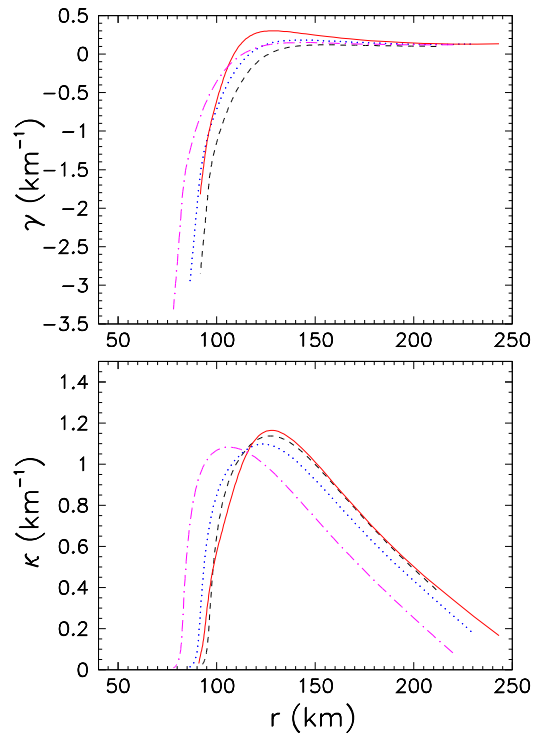}
\caption{Inverted mass hierarchy, case $\cal A$. Multi-angle eigenvalues $\gamma(r)$ and $\kappa(r)$. 
The different curves correspond to $\beta_e=\beta_x=0.0$ (dashed curves), $\beta_e=1.0, \beta_x=1.5$
 (dotted curves), $\beta_e=1.0, \beta_x=3.0$ (continuous curves), $\beta_e=3.0, \beta_x=3.0$
(dash-dotted curves).  
\label{fig3}} 
\end{figure}

\begin{figure}  
\includegraphics[angle=0,width=0.8\columnwidth]{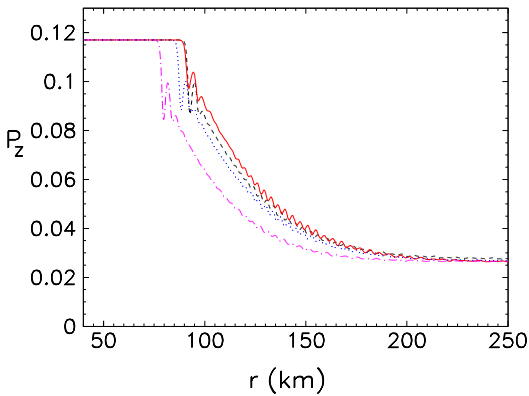}
\caption{Inverted mass hierarchy, case $\cal A$. Radial evolution of the integrated $z$-component
of the neutrino polarization vector $P_z$.
The different curves correspond to $\beta_e=\beta_x=0.0$ (dashed curve), $\beta_e=1.0, \beta_x=1.5$
 (dotted curve), $\beta_e=1.0, \beta_x=3.0$ (continuous curve), $\beta_e=3.0, \beta_x=3.0$
(dash-dotted curve).  
\label{fig4}} 
\end{figure}

Hence, we present only the results for the IH case where the instability is expected to occur. 
Let us discuss first the half-isotropic case  $\beta_e=\beta_x=0.0$. 
In the case of a single-crossing spectrum, if the consistency equations [Eqs.~(\ref{eq:JnKn})] admit a solution,
this is a unique $(\gamma, \kappa)$  for a given $\mu$, or equivalently radial location in the SN~\cite{Banerjee:2011fj}. 
In Figure~\ref{fig3} we show the  $(\gamma, \kappa)$  solution as a function of $r$  (dashed curves for the 
half-isotropic case). 
The onset of the flavor conversions occurs when $\kappa$ starts to grow, i.e. at $r \simeq 90$~km. 
In order to compare this prediction with the numerical results of the integration of the equations of motion
[Eqs.~(\ref{eq:eom1})-(\ref{eq:eom2})], we represent in Figure~\ref{fig4} the integrated value
of the $z-$component of the $\nu$ polarization vector $P_z$, that is related to the flavor content of the ensemble. 
The onset of the flavor conversions for the half-isotropic case (dashed curve) is in good agreement with expectations from the stability analysis. 

We pass now to analyze the cases with other neutrino angular distributions presented in Sec.~IIIA. 
Figure~\ref{fig3} shows that different non-trivial cases do not lead to major changes in the behavior
of the solution $(\gamma, \kappa)$ of Eq.~(\ref{eq:consit}). In particular, no enhancement in the value of
$\kappa$ is observable. This result is consistent with the numerical simulations
of the flavor evolution (see Fig.~\ref{fig4}) that does not show any significant change in the non-trivial cases with 
respect to the half-isotropic one. 
The stability analysis confirms that in absence of crossing points in the angular spectra, no new multi-angle instability is triggered by the non-trivial angular distributions {\it alone}. 
The only observable consequence of the non-trivial angular distributions is a shift  in the onset in the 
flavor conversions toward smaller $r$. 
 Since  forward-peaked distributions  receive mostly 
 contributions from small values of $u$, the resonance condition in Eq.~(\ref{eq:res}) can be satisfied at larger $\mu$,
  i.e. at smaller $r$ with respect to the half-isotropic case. This is consistent with the expectation
that the $\nu$-$\nu$ strength is weaker for forward-peaked distributions, making the  system to become unstable  
 earlier than in the half-isotropic case. 
In particular, we find as onset of the flavor conversions  $r= 87$~km for the case with
$\beta_e=1.0, \beta_x=1.5$, $r= 88$~km, for the case with
$\beta_e=1.0, \beta_x=3.0$, and $r= 78$~km for $\beta_e=3.0, \beta_x=3.0$, consistently with what shown in 
Figure~\ref{fig4}.
Finally, we also checked that no effect is triggered in NH by the presence of non-trivial
angular distributions. 
 
\subsection{Case ${\cal C}$}

The $g_{\omega}$  spectrum for the case  ${\cal C}$   has been widely studied in the bulb model 
as  representative of energy spectra with a flux ordering
$N_{\nu_x} \gtrsim N_{\nu_e} \gtrsim N_{\bar\nu_e}$ and energy spectra 
 with multiple crossing points  leading to multiple 
splits around these points (see, e.g.,~\cite{Dasgupta:2009mg,Duan:2010bf,Mirizzi:2010uz}).  
Since we expect flavor instabilities in both the mass hierarchies, we discuss 
separately these two cases.
 
\subsubsection{Inverted mass hierarchy}

\begin{figure}[!t]  
\includegraphics[angle=0,width=0.8\columnwidth]{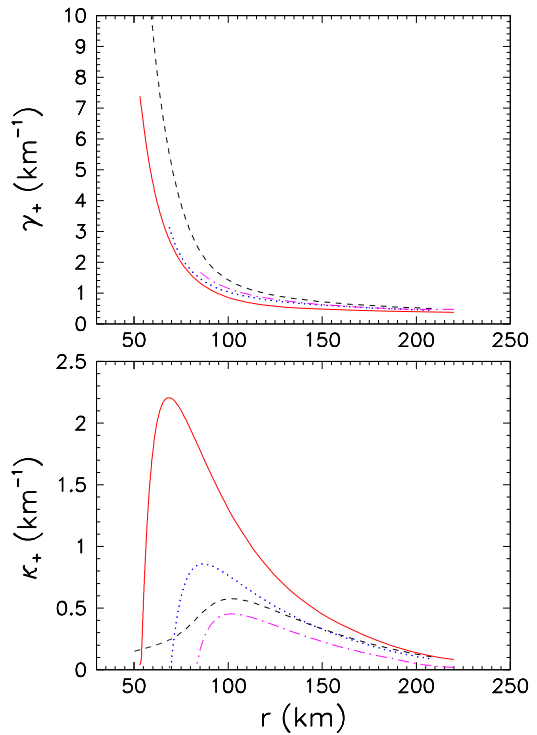}
 \caption{Inverted mass hierarchy, case $\cal C$. Multi-angle eigenvalues $\gamma_+(r)$ and $\kappa_+(r)$.   The different curves correspond to $\beta_e=\beta_x=0.0$ (dashed curves), $\beta_e=1.0, \beta_x=1.5$
 (dotted curves), $\beta_e=1.0, \beta_x=3.0$ (continuous curves), $\beta_e=3.0, \beta_x=3.0$
(dash-dotted curves).
\label{fig5}} 
\end{figure}  

\begin{figure}[!t]   
\includegraphics[angle=0,width=0.8\columnwidth]{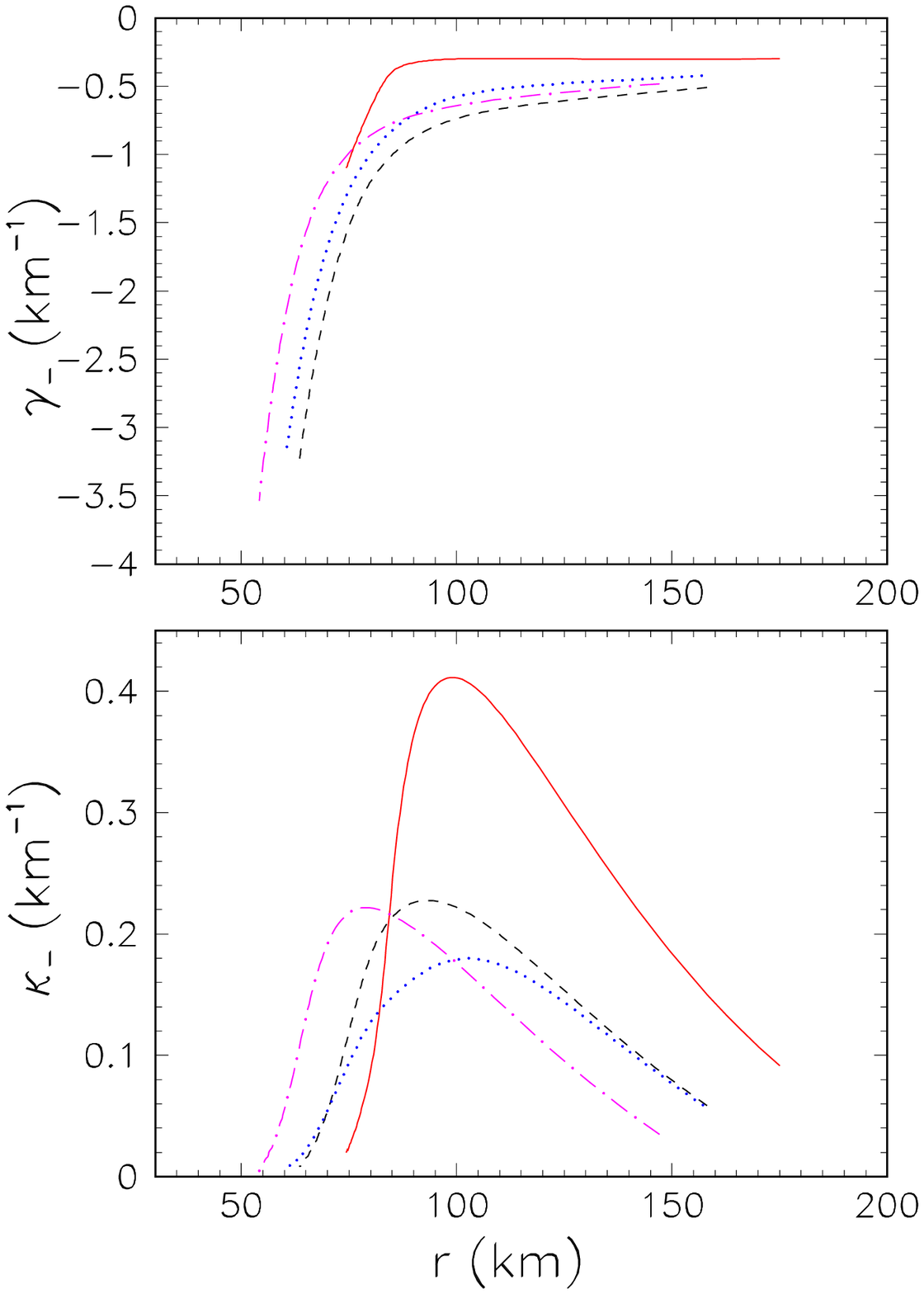}
\caption{Inverted mass hierarchy, case $\cal C$. Multi-angle eigenvalues $\gamma_-(r)$ and $\kappa_-(r)$. The different curves correspond to $\beta_e=\beta_x=0.0$ (dashed curves), $\beta_e=1.0, \beta_x=1.5$
 (dotted curves), $\beta_e=1.0, \beta_x=3.0$ (continuous curves), $\beta_e=3.0, \beta_x=3.0$
(dash-dotted curves).
\label{fig6}} 
\end{figure}  

\begin{figure}  
\includegraphics[angle=0,width=0.8\columnwidth]{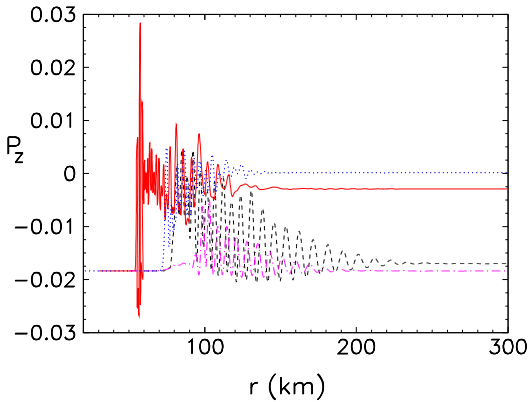}
\caption{Inverted mass hierarchy, case $\cal C$. Radial evolution of the integrated $z$-component
of the neutrino polarization vector $P_z$.
The different curves correspond to $\beta_e=\beta_x=0.0$ (dashed curve), $\beta_e=1.0, \beta_x=1.5$
 (dotted curve), $\beta_e=1.0, \beta_x=3.0$ (continuous curve), $\beta_e=3.0, \beta_x=3.0$
(dash-dotted curve).  
\label{fig7}} 
\end{figure}  

As discussed in~\cite{Banerjee:2011fj}, with spectra presenting three crossing points the consistency equations   
[Eq.~(\ref{eq:JnKn})] admit a pair of solutions $(\gamma_{+}, \kappa_{+})$ and $(\gamma_{-}, \kappa_{-})$ (of course, if they admit a solution at all) for each 
value of $\mu$, corresponding
to negative and positive values of $\gamma$ respectively. 
We show in Fig.~\ref{fig5} and \ref{fig6}  the functions  $(\gamma_{+}(r), \kappa_{+}(r))$ and $(\gamma_{-}(r), \kappa_{-}(r))$, respectively,
for the different angular distributions discussed before. 
A comparison between these two Figures reveals that $\kappa_{-}$ is always smaller than 
$\kappa_{+}$, meaning that the instability associated with $\kappa_{-}$ will be always sub-leading.
 Therefore, in the following it suffices to discuss only the $(\gamma_{+}, \kappa_{+})$ solution.
In the  half-isotropic case $\beta_e=\beta_x=0$ (dashed curve) the $\kappa_+$ function presents
a hump peaked around $r \simeq 100$~km and connected with a long tail at smaller $r$ around $r \simeq 75$~km. 
This tail indicates that the system is in principle  always unstable (also at very large values of $\mu$). However, at
large $\mu$ and  $\gamma_{+}$, the unstable frequency modes 
satisfying the resonance condition [Eq.~(\ref{eq:res})]
are in the infrared region (i.e. $|\omega| > 4$), 
where the $g_\omega$ spectrum is strongly suppressed. Therefore, they have no 
impact for the flavor conversions. 
The onset of the flavor conversions is then given by the connection between the hump and the tail
 at  $r=75$~km. This result is  in agreement with what shown in the numerical calculation
of $P_z$, shown in Fig.~\ref{fig7} (dashed curve).

When considering non-trivial angular distributions, the long tail in the
$\kappa_+$  function observed in the half-isotropic case at small $r$ now disappears. 
However, the presence of a non-isotropic angular distribution is not enough to produce an enhancement in the value of
 $\kappa_+$. 
Conversely, in the flavor blind case $\beta_e=\beta_x=3.0$ (dot-dashed curve) the instability is suppressed with respect to what 
has been seen in the half-isotropic case. 
 Indeed, around  the resonance energies [Eq.~(\ref{eq:res})],  the integrals $J_n$ and $K_n$
 in Eq.~(\ref{eq:JnKn})
are proportional to  $\int du\, d\omega\, g_{\omega,u}\, u^n\, \kappa^{-2}$. 
Since with forward-peaked distributions these integral receive mostly contributions from small $u$,
in order to satisfy the consistency equations [Eq.~(\ref{eq:JnKn})] at a given $\mu$,
the corresponding $\kappa_+$ has to  be smaller than in the half-isotropic
 case.
 In this case, flavor conversions start at $r=80$~km.

 A significant enhancement of the multi-angle instability occurs when $\nu$ angular spectra exhibit crossing points 
 in $u$, as can be seen in the case  with $\beta_e=1.0, \beta_x=1.5$ (dotted curve)  and even more in the one
 with $\beta_e=1.0, \beta_x=3.0$ (continuous curve).
 In both  cases the peak in $\kappa_+$ increases with respect to the half-isotropic case and 
 also the hump broadens toward smaller $r$.
Since $\kappa_+$ reaches a higher peak value, 
the width of the Lorentzian around an unstable frequency mode for a given angle $u$ would be broad,
implying a speed-up in the transitions.  
 In particular, in the case of $\beta_e=1.0, \beta_x=1.5$ flavor conversions  start
 around  $r\simeq 70$~km, while in the case with  $\beta_e=1.0, \beta_x=3.0$
 around  
$r \simeq 55$~km, in agreement
with the numerical results shown in Fig.~\ref{fig7}.

\begin{figure}[!t]  
\includegraphics[angle=0,width=0.8\columnwidth]{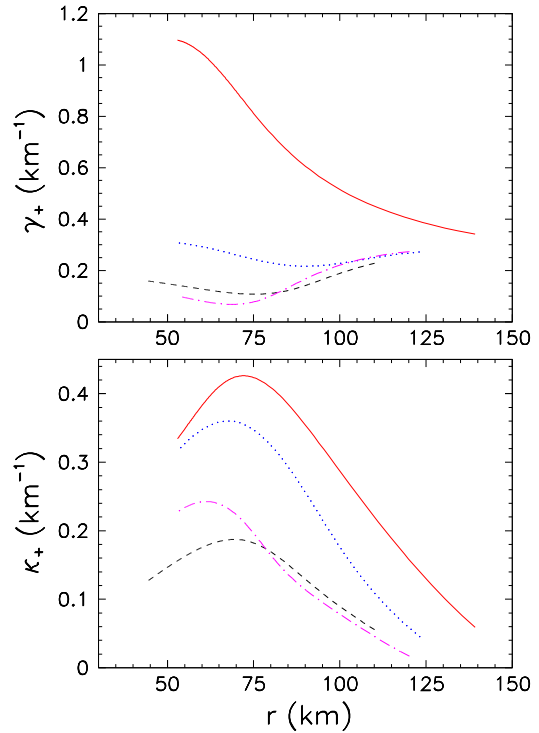}
 \caption{Normal mass hierarchy, case $\cal C$. Multi-angle eigenvalues $\gamma_+(r)$ and $\kappa_+(r)$. The different curves correspond to $\beta_e=\beta_x=0.0$ (dashed curves), $\beta_e=1.0, \beta_x=1.5$
 (dotted curves), $\beta_e=1.0, \beta_x=3.0$ (continuous curves), $\beta_e=3.0, \beta_x=3.0$
(dash-dotted curves).
\label{fig8}} 
\end{figure}  

\begin{figure}[!t]  
\includegraphics[angle=0,width=0.8\columnwidth]{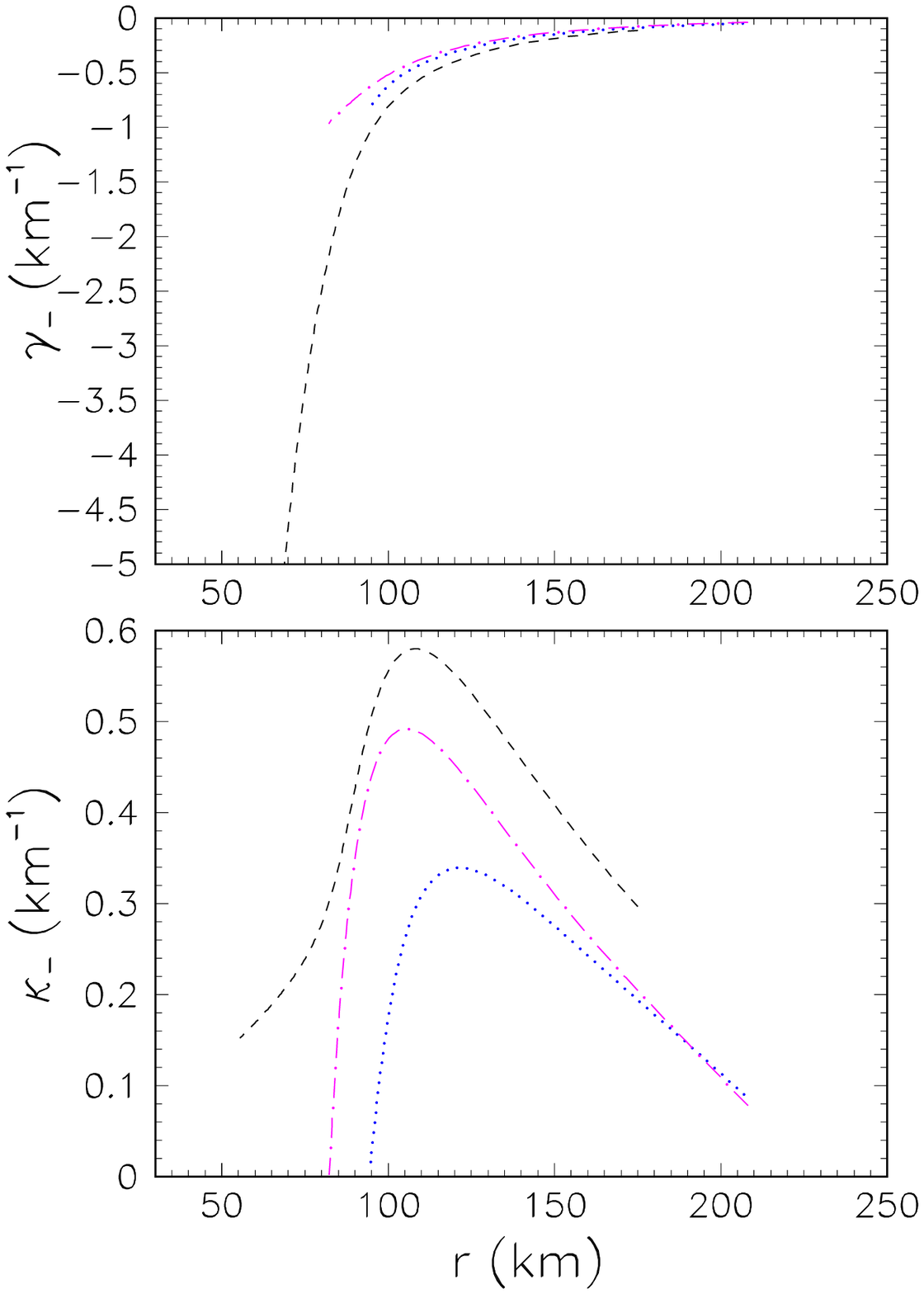} 
\caption{Normal mass hierarchy, case $\cal C$. Multi-angle eigenvalues $\gamma_-(r)$ and $\kappa_-(r)$. The different curves correspond to $\beta_e=\beta_x=0.0$ (dashed curves), $\beta_e=1.0, \beta_x=1.5$
 (dotted curves), $\beta_e=1.0, \beta_x=3.0$ (continuous curves), $\beta_e=3.0, \beta_x=3.0$
(dash-dotted curves).
\label{fig9}} 
\end{figure}  

\begin{figure}  
\includegraphics[angle=0,width=0.8\columnwidth]{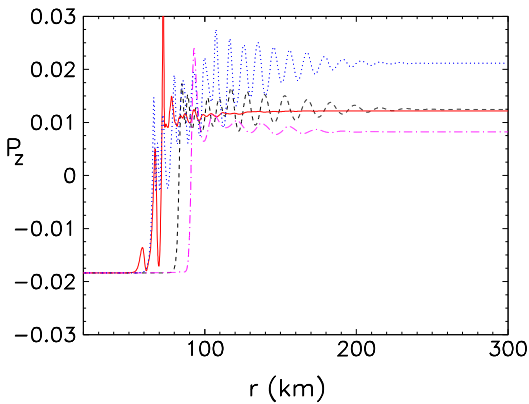}
\caption{Normal mass hierarchy, case  $\cal C$.
Radial evolution of the integrated $z$-component
of the neutrino polarization vector $P_z$.
The different curves correspond to $\beta_e=\beta_x=0.0$ (dashed curve), $\beta_e=1.0, \beta_x=1.5$
 (dotted curve), $\beta_e=1.0, \beta_x=3.0$ (continuous curve), $\beta_e=3.0, \beta_x=3.0$
(dash-dotted curve).  
\label{fig10}} 
\end{figure}  

In conclusion, we remark that  SN $\nu$ spectra with multiple crossing points in the energy domain can 
exhibit flavor conversions at very large $\mu$ in a single-angle scheme~\cite{Raffelt:2008hr}. Multi-angle effects
in a half-isotropic situation suppress the instability  
at  large neutrino densities~\cite{Raffelt:2008hr,Duan:2010bf}. However, the
stability analysis and our numerical simulations tell us that this stabilization effect in not
a generic situation. Indeed, allowing for crossing points also in the angular spectra, flavor conversions
can start much earlier than in the half-isotropic case.

\subsubsection{Normal mass hierarchy}

We finally pass to analyze the normal mass hierarchy case. 
Also here, we expect that the consistency equations 
 [Eq.~(\ref{eq:JnKn})]  admit a pair of solutions
 $(\gamma_{+}, \kappa_{+})$ and $(\gamma_{-}, \kappa_{-})$ for each 
value of $\mu$, that we represent in function of $r$ in Fig.~\ref{fig8} and \ref{fig9} respectively. 
The two imaginary parts of the eigenvalues, $\kappa_+$ and  $\kappa_-$ can assume comparable values,
differently from the IH case.
Whether $\kappa_+$  or $\kappa_-$ plays the dominant role in triggering the flavor conversions will depend on the specific case.
Starting with the half-isotropic case ($\beta_e=\beta_x=0.0$, dashed curves) we see that
both the functions $\kappa_+$  and $\kappa_-$ present long tails extending at small $r$.
However, 
 $\kappa_+$ is always smaller than $0.2$, 
 while $\kappa_-$
reaches a peak value $\sim 0.6$ at $r \simeq 105$~km.
Therefore, the onset of the flavor conversions will be associated to  
$\kappa_-$, and as usual is determined by the connection of the hump with the tail of the function,
i.e. at  $r\simeq 80$~km, in agreement with the numerical evolution
of $P_z$ shown in Fig.~\ref{fig10} (dashed curve for the $\beta_e=\beta_x=0.0$ case).  
In the flavor-blind case with $\beta_e=\beta_x=3.0$ (dash-dotted curves) the dominant instability is again the one associated with $\kappa_-$
that can reach as peak value $\sim 0.5$ at $r \simeq 105$~km. 
We note that  in this case  $\kappa_-(r)$ has no tail, terminating abruptly
at $r\simeq 80$ km, while $\kappa_+(r)$ extends toward lower $r$ reaching a almost constant value 
$\sim 0.2$. 
 The predicted onset 
of the flavor conversions can then be located at  $r\simeq 80$ km.
We see the corresponding numerical evolution in 
 Fig.~\ref{fig10} (dash-dotted curve)  starts at a slight larger radius, i.e.  $r\simeq 90$~km.
 This delay  is probably due to the time needed by the instability to grow significantly after it 
 starts. This is related to the value of $\kappa_{-}$ that is relatively small in this case.
 In the case  $\beta_e=1.0, \beta_x=1.5$ (dotted curves)  $\kappa_-$ is further reduced with respect to the previous cases, while
  $\kappa_+$ is enhanced. Both the functions have a  similar peak $\sim 0.35$, but 
$\kappa_+$ has a longer tail with a value close to the peak one, while $\kappa_-$  abruptly drops at zero for
$r \simeq 95$~km. 
Therefore, in this case the onset is determined by  $\kappa_+$. Since the instable frequencies, determined
by the resonance condition [Eq.~(\ref{eq:res})] would be in the infrared range (i.e. $\omega>4$) for $\mu>50$,
the onset of relevant flavor conversions  corresponds to this value of $\mu$, i.e.  $r\simeq 65$ km, as visible 
in the numerical evolution  (dotted curve). 
Finally, in the case   $\beta_e=1.0, \beta_x=3.0$ (continuous curve), the instability associated with $\kappa_-$ disappears, while the one 
associated with $\kappa_+$ is further enhanced, reaching as peak value $\sim 0.4$. 
This time the resonance condition  reaches infrared frequency modes for $\mu \gtrsim 27$, 
corresponding to an onset of the flavor conversions at $r\lesssim77$ km. 
Note that in this case the numerical evolution (continuous curve) the $P_z$ starts to vary earlier, 
around  $r=55$ km, but it rises significantly only at $r\gtrsim 70$ km, as predicted by the stability analysis. This can be
interpreted as follows:
in the  latter two cases, since the $\kappa_{-}$ function presents a rather flat behavior in $\mu$,  the determination of the onset of  the flavor conversions in less accurate.   
Indeed, also the numerical evolution in Fig.~\ref{fig10} shows that the $P_z$-component presents a slower rise time at the onset of the conversions.

\section{Conclusions} \label{conclusions}

Self-induced flavor conversions for SN neutrinos are associated to instabilities in the flavor space.
Recently, a linearized stability analysis of the SN $\nu$ equations of motion has been proposed as diagnostic tool to track the emergence of these  instabilities~\cite{Banerjee:2011fj}. 
In our work we have applied this technique to study the multi-angle instability associated with
non-trivial neutrino angular distributions, extending the results presented in~\cite{Mirizzi:2011tu}
to different SN $\nu$ flux orderings and  $\nu$ mass hierarchies. We confirm that
enhanced instabilities can occur,  when the angular spectra present crossing points.
Also, the onset of the flavor conversions typically shifts toward small radii, with respect to the
half-isotropic case. 

We cannot but emphasize that neutrino angular distributions should be taken into account
in supernova neutrino phenomenological studies. For example,  as pointed out in~\cite{Mirizzi:2011tu}, 
one can expect a smearing of the splitting features observed in the half-isotropic case, with resulting $\nu$ fluxes showing less significant spectral differences. This tendency toward spectral equalization would
challenge the detection of further oscillation signatures, like the ones associated with the
Earth crossing of SN $\nu$'s (see, e.g.,~\cite{Lunardini:2001pb,Borriello:2012zc}). 
On the other hand, a sufficiently low-radii onset on the flavor conversions might imply an interesting impact on the $r$-process nucleosynthesis in SNe~\cite{Duan:2010af}. Eventually, if $\nu$ oscillations develop too close to the neutrinosphere, they might invalidate the $\nu$ transport paradigm in SNe that ignores  $\nu$ conversions.

It is clear that future, state-of-the-art predictions of the neutrino flavor evolution in SNe would require as accurate as possible input from hydrodynamical simulations, providing the flavor, angular, energy and time structure of the fluxes. 
Here we showed that a preliminary flavor stability analysis presents some key advantages over a brute force numerical integration. While such a perturbative analysis
cannot be used to deduce the eventual fate of the neutrino ensemble in flavor space, it can be useful to identify 
the situations where angular distributions  play a role in the flavor evolution.  Once these cases are identified, it would 
be mandatory to perform large scale numerical simulations of the complete flavor evolution in order 
to determine the final oscillated neutrino spectra. With current level of sophistication in SN simulations, we believe that
such studies will be mandatory to achieve a better characterization of the non-linear neutrino flavor evolution during a 
stellar gravitational collapse and a more realistic description of the 
neutrino signal from a future galactic SN.

\section*{Acknowledgments} 
We acknowledge Nicolas Bourbaki for useful discussions.
We thank Georg Raffelt and Irene Tamborra for reading the manuscript and for interesting comments on it. 
The work of  A.M. was supported by the German Science Foundation (DFG) within the Collaborative Research Center 676 ``Particles, Strings and the Early Universe''. 


\end{document}